\documentclass[%
 reprint,
 amsmath,amssymb,
 aps,
]{revtex4-2}
\usepackage{caption}

\usepackage{tabularx}
\usepackage{graphicx}% Include figure files
\usepackage{dcolumn}% Align table columns on decimal point
\usepackage{bm}% bold math
\usepackage{braket}
\begin{document}

\preprint{APS/123-QED}

\title{Enhanced Critical Field of Superconductivity at an Oxide Interface}
\author{Athby H. Al-Tawhid}
\altaffiliation{These authors contributed equally to this work}
\affiliation{Department of Materials Science and Engineering, North Carolina State University, Raleigh, NC 27265 USA} 
\author{Samuel J. Poage}
\altaffiliation{These authors contributed equally to this work}
\affiliation{Department of Materials Science and Engineering, North Carolina State University, Raleigh, NC 27265 USA} 
\author{Salva Salmani-Rezaie}
\affiliation{School of Applied and Engineering Physics, Cornell University, Ithaca, New York 14853, USA}
\affiliation{Kavli Institute at Cornell for Nanoscale Science, Cornell University, Ithaca, New York 14853, USA}
\author{Antonio Gonzalez}
\affiliation{Department of Materials Science and Engineering, North Carolina State University, Raleigh, NC 27265 USA}
\author{Shalinee Chikara}
\affiliation{National High Magnetic Field Laboratory, Tallahassee, Florida 32310, USA}
\author{David A. Muller}
\affiliation{School of Applied and Engineering Physics, Cornell University, Ithaca, New York 14853, USA}
\affiliation{Kavli Institute at Cornell for Nanoscale Science, Cornell University, Ithaca, New York 14853, USA}
\author{Divine P. Kumah}
\affiliation{Department of Physics, North Carolina State University, Raleigh, North Carolina 27695, USA}
\author{Maria N. Gastiasoro}
\affiliation{Donostia International Physics Center, 20018 Donostia-San Sebastian, Spain}
\author{Jos\'e Lorenzana}
\affiliation{ISC-CNR and Department of Physics, Sapienza University of Rome, Piazzale Aldo Moro 2, 00185, Rome, Italy}
\author{Kaveh Ahadi}
 \email{kahadi@ncsu.edu}
\affiliation{Department of Materials Science and Engineering, North Carolina State University, Raleigh, NC 27265 USA}
\affiliation{Department of Physics, North Carolina State University, Raleigh, North Carolina 27695, USA}

\date{\today}% It is always \today, today,
             %  but any date may be explicitly specified

\begin{abstract}

The nature of superconductivity and its interplay with strong spin-orbit coupling at the KTaO$_{3}$(111) interfaces remains a subject of debate. 
%KTaO$_{3}$ is an incipient ferroelectric with strong spin-orbit coupling. 
To address this problem, we grew epitaxial LaMnO$_{3}$/KTaO$_{3}$(111) heterostructures. We show that superconductivity is robust against the in-plane magnetic field, with the critical field of superconductivity reaching $\sim 25$ T in optimally doped heterostructures. The superconducting order parameter is highly sensitive to carrier density. We argue that spin-orbit coupling drives the formation of anomalous quasiparticles with vanishing magnetic moment, providing the condensate significant immunity against magnetic fields beyond the Pauli paramagnetic limit. These results offer design opportunities for superconductors with extreme resilience against magnetic field.

\end{abstract}

%\keywords{Suggested keywords}%Use show keys class option if keyword
                              %display desired
\maketitle

%\tableofcontents

\section{Introduction}
Spin-orbit coupling (SOC) in two-dimensional (2D) superconductors gives rise to robust superconductivity against the applied magnetic fields \cite{saito2016highly, zhang2021ising}. For example, spin-orbit induced Zeeman spin splitting competes with in-plane magnetic field in Ising superconductors, stabilizing a highly resilient superconductivity \cite{lu2015evidence, xi2016ising, falson2020type, rhodes2021enhanced}. Furthermore, Rashba SOC, creating an in-plane helical spin texture, enhances the superconducting critical field beyond the Pauli paramagnetic limit (${H_{P}}$) \cite{gor2001superconducting}. Rashba SOC enhancement of critical field, however, is limited to $\sqrt{2}{H_{P}}$. 
An important effect of SOC is the generation of states with entangled spin and orbital degrees of freedom, which may lead to single particle states with vanishing magnetic moments. While the ordering of such ``non-magnetic" electrons in Mott insulators has been studied \cite{Ishikawa2019}, itinerant states and the consequences for superconductivity remain unexplored. 

%Large 
%In this work, we argue that large SOC could also create quasiparticles with negligible effective magnetic moment and, accordingly, small coupling to applied magnetic field. Superconductivity emerging from these quasiparticles survives anomalously large magnetic fields. 

KTaO$_{3}$ is an incipient ferroelectric which remains cubic down to low temperature \cite{shirane1967temperature, rowley2014ferroelectric}. A 2D superconductivity was recently discovered in  (111) interfaces \cite{liu2021two}. 
Figure 1(a) shows the schematic crystal structure of the KTaO$_{3}$ (111) surface. Itinerant electrons occupy the tantalum 5$d$ derived $t_{2g}$ states in electron-doped samples \cite{himmetoglu2016transport}.

From a magnetic point of view, quasiparticles in KTaO$_{3}$ can be expected to be highly anomalous. Analogous to photons, which can carry polarization and angular momentum\cite{Bliokh2015}, 
electronic quasiparticles in KTaO$_{3}$
 carry not only spin $\hat{\bm s}$  but also orbital angular momentum, $\hat{\bm l}$. This is due to the strong SOC in the constituent 5$d$ $t_{2g}$ orbitals of Ta.
 Both $\hat{\bm l}$ and $\hat{\bm s}$
contribute to the quasiparticle magnetic moment $\hat{\bm m} =-\mu_B (\hat{\bm l} +2 \hat{\bm s})$. 
Furthermore, $t_{2g}$ states can be mapped \cite{Sugano1970,Stamokostas2018} into $p$ states with a fictitious inverted orbital angular momentum $\hat{\bm l}_p=-\hat{\bm l}$ (partially quenched from $l=2$ to $l_p=1$), and an \emph{effective} total angular momentum $\hat{\bm j}=(\hat{\bm l}_p + \hat{\bm s})$. 
%A simplified picture of the electronic structure can be obtained considering only the two uppermost Ta layers\cite{xiao2011interface,villar2021}. 
Figure~1(b) illustrates the lowest energy bands of a simplified electronic model of the KTaO$_{3}$ (111) surface.
%this model with Rahsba split bands similar to first principle computations\cite{bruno2019band}.
To a good approximation, the states near $\Gamma$ have an effective angular momentum $j=3/2$ in which $\hat{\bm l}_p$ and $\hat{\bm s}$ are parallel, implying a magnetic moment $\hat{\bm m} =-\mu_B (-\hat{\bm l} _p+2 \hat{\bm s})\approx 0$. 
Figure 1(c) shows the texture of the magnetic moment $\hat{\bm m}$ for Fermi surface quasiparticles illustrating its strong reduction even when the mixing between states with different $j$ is allowed. This remarkable property suggests a small coupling of quasiparticles with an in-plane magnetic field and therefore a strong resilience of superconductivity.

Here, we report on emergence of a 2D superconductivity with enhanced critical field and carrier density dependence in molecular beam epitaxy (MBE) grown KTaO$_{3}$ (111) epitaxial interfaces. The in-plane critical field ($H_{c2}$) exceeds the Pauli limit and reaches $\sim 25$ T in optimally doped films. $H_{c2}$ is highly sensitive to carrier density and magnetic field orientation. To understand the resilience of superconductivity, we computed the critical field in a two-layer model of a  KTaO$_{3}$ (111) interface.

\section{Results}
Epitaxial LaMnO$_{3}$/KTaO$_{3}$(111) heterostructures were grown in an oxide MBE with various 2D electron densities. Figure S1 shows the reflection high-energy electron diffraction (RHEED) streaks, suggesting two-dimensional growth of LaMnO$_{3}$. High-angle annular dark-field scanning transmission electron microscope (HAADF-STEM) was used to investigate the grown heterostructures. Figure 2(a) shows the HAADF-STEM image of LaMnO$_{3}$ grown on a KTaO$_{3}$(111) substrate. The HAADF-STEM image confirms epitaxial growth of LaMnO$_{3}$ on KTaO$_{3}$(111). The zoomed-in region shows a continuous bright contrast at the interface which could be due to the steps at the (111) interface \cite{raghavan2015two}. Figure S2 shows the Hall carrier mobility, measured for samples with different carrier densities ($n_{s}=3.5\times10^{13}$ $cm^{-2}$, $5.1\times10^{13}$ $cm^{-2}$, $8.8\times10^{13}$ $cm^{-2}$, and $9.2\times10^{13}$ $cm^{-2}$). The measured 2D carrier densities fall in the range that a superconducting transition is expected \cite{liu2023tunable}. Figure S2 demonstrates the carrier mobility at different temperatures. The carrier mobility at liquid helium temperature reaches its maximum, $\sim300$ $cm^{2}/v.s$ for $n_{s}=3.5\times10^{13}$ $cm^{-2}$, in line with high quality KTaO$_{3}$(111) 2D electron systems \cite{liu2021two}. The carrier mobility is systematically suppressed with increasing carrier density.

Figure 2(c) shows the resistance with temperature (300-0.1 K), measured along the [11$\bar2$] direction. All films exhibit a metallic behavior, $dR/dT>0$, followed by an abrupt superconducting transition. Similar heterostructures, grown on sapphire substrate instead of KTaO$_{3}$, show insulating behavior, suggesting the transport occurs at the KTaO$_{3}$ side of the interface. The critical temperature of superconductivity ($T_{c}$ defined at 0.1$R_{n}$) is 0.37 K, 0.82 K, 1.91 K, and 1.48 K for $n_{s}=3.5\times10^{13}$ $cm^{-2}$, $5.1\times10^{13}$ $cm^{-2}$, $8.5\times10^{13}$ $cm^{-2}$, and $9.2\times10^{13}$ $cm^{-2}$, respectively. These transition temperatures are comparable to those found at EuO/KTaO$_{3}$(111) and LaAlO$_{3}$/KTaO$_{3}$(111) interfaces \cite{liu2021two, chen2021electric}. We estimate the BCS superconducting gap ($\Delta \approx 1.76 k_B T_c$) to be 60 $\mu eV$, 120 $\mu eV$, 290 $\mu eV$, and 220 $\mu eV$ for $n_{s}=3.5\times10^{13}$ $cm^{-2}$, $5.1\times10^{13}$ $cm^{-2}$, $8.5\times10^{13}$ $cm^{-2}$, and $9.2\times10^{13}$ $cm^{-2}$, respectively

We further study the superconducting transition by measuring the longitudinal transport with magnetic field. Figure 3 shows the longitudinal magnetoresistance at different angles between the current and magnetic field. We find that the in-plane critical fields ($H_{c2}$ defined at 0.9$R_{n}$) are $23.1$ T , $24.7$ T, $5.31$ T, and $1.83$ T for $n_{s}=9.2\times10^{13}$ $cm^{-2}$, $8.5\times10^{13}$ $cm^{-2}$, $5.1\times10^{13}$ $cm^{-2}$, and $3.5\times10^{13}$ $cm^{-2}$, respectively. Figure 4(a) compares the critical temperatures and critical fields in heterostructures with various carrier densities. The critical temperature increases somewhat linearly with carrier density while critical field is much more sensitive to density of charge carriers. Additionally, there is a significant asymmetry between out-of-plane and in-plane $H_{c2}$ results, an effect commonly observed in 2D superconductors \cite{saito2016highly}. This asymmetry, however, varies with carrier density and is illustrated in Figure S3. We also note the in-plane critical fields exceed the Pauli limit, which we estimate to be $H_p \approx \frac{\Delta}{\sqrt{2}\mu_B} $, assuming a $g$-factor of 2 and weakly coupled superconductivity. Table I shows the $\frac{H_{c2, \parallel}} {H_{p}}$ grows with increasing carrier density.    

Figure 4(b) shows $H_{c2}$ (defined at 0.9$R_{n}$) follows the equation 1, first derived by Tinkham for two-dimensional superconductors \cite{tinkham1963effect}. Figure S4 exhibits the $H_{c2}$, defined at various $R_{n}$ fractions also follow a similar trend. 
\begin{equation}
\left|
%\abs{\lvert}
\frac{H_{c2}(\theta)\sin\theta}{H_{c2, \perp}}%{\rvert}
\right|+\left(\frac{H_{c2}(\theta)\cos\theta}{H_{c2, \parallel}}\right)^2=1,
\label{eq1}
\end{equation}
 Figure S5 demonstrates the in-plane and out-of-plane critical fields with temperature, illustrating a monotonic increase of $H_{c2}$ as we cool down the sample.

\begin{table}[h!]\label{Table}
    \caption{Values for $T_{c}$ (defined at 0.1$R_{n}$), $H_{C2\parallel and \perp}$ (defined at 0.9$R_{n}$), $\Delta$ ($1.76 k_B T_c$), $\frac{H_{C2\parallel}} {H_{C2\perp}}$, and $\frac{H_{C2\parallel}} {H_{P}}$ ($H_p \approx \frac{\Delta}{\sqrt{2}\mu_B} $) of KTaO$_{3}$(111) 2D electron systems with various carrier densities (measured at base temperature).}
    \begin{center}
\resizebox{0.5\textwidth}{!}
{
\begin{tabular}{|l l l l l l l|} 
 \hline
 $n_{s}(cm^{-2})$ & $T_{c}$(K) & $H_{C2\parallel}$(T) & $H_{C2\bot}$(T) & $\Delta$($\mu$eV) & $\frac{H_{C2\parallel}} {H_{C2\perp}}$ & $\frac{H_{C2\parallel}} {H_{P}}$ \\ [0.5ex] 
 \hline\hline
 $3.5\times10^{13}$ & 0.37 & 1.83 & 0.175 & 60 & 10.5 & 2.7 \\ 
 $5.1\times10^{13}$ & 0.82 & 5.31 & 0.28 & 120 & 19.0 & 3.5 \\ 
 $8.5\times10^{13}$ & 1.91 & 24.7 & 1.71 & 290 & 14.4 & 7.0 \\
 $9.2\times10^{13}$ & 1.48 & 23.1 & 1.23 & 220 & 18.8 & 8.4 \\  
 \hline
\end{tabular}
}
 \end{center}
    \end{table}

\section{Discussion}

Table I summarizes the measured results in KTaO$_{3}$(111) interfaces with various carrier densities. The resilience of superconducting state against in-plane magnetic field and violation of Pauli limit could be explained by the emergence of the Fulde–Ferrell–Larkin–Ovchinnikov (FFLO) state \cite{fulde1964superconductivity}. In the presence of finite Rashba-type SOC, FFLO state may arise. The FFLO state, however, emerges in clean regime ($l_{mfp}>>\zeta_\parallel$). Here, the in-plane coherence length is $\zeta_\parallel\approx 43$ nm, $\zeta_\parallel=\sqrt{\Phi_0/2\pi H_{c2,\bot}}$, where $\Phi_0$ is flux quantum, for $n_{s}=3.5\times10^{13}$. The mean free path of charge carriers at 3 K is $\approx 25$ nm (supplementary materials), suggesting the system is not in the clean limit. The samples with higher carrier densities demonstrate smaller mean free path. 

Rashba-type spin splitting was reported at the KTaO$_{3}$(111) surface \cite{bruno2019band}. A Rashba SOC acting on spin-1/2 quasiparticles which carry no orbital angular momentum is expected to enhance the Pauli limit by a factor of $\sqrt{2}$ \cite{gor2001superconducting}. Here, the $\frac{H_{C2, \parallel}} {H_{P}}$ is 2.7 for $n_{s}=3.5\times10^{13}$ $cm^{-2}$ and increases with carrier density to 8.4 for $n_{s}=9.2\times10^{13}$ $cm^{-2}$ much beyond the prediction of a simple Rashba model.

To understand the resilience of superconductivity against the in-plane magnetic field, we  examine the nature of the electronic quasiparticles at the KTaO$_3$ (111) surface. The charge accumulation at the KTaO$_3$/LaMnO$_3$ interface creates a negative band-bending potential, which can quantize the Ta-$t{2g}$ conduction bands into distinct subbands. A (111) bilayer model with three $5d$ $t_{2g}$ Wannier orbitals per Ta site ($yz,zy,xy$) and two layers forming a buckled honeycomb plane~\cite{xiao2011interface, villar2021} captures many of the key features of the band structure measured by ARPES~\cite{bruno2019band}. This minimal model includes an inter-layer nearest-neighbor hopping between Ta-I and Ta-II atoms
in Fig.~1(a), a small trigonal crystal field, the strong local atomic SOC for the $5d$ $t_{2g}$ orbitals and a Rashba term due to broken inversion symmetry at the interface (see Materials and Methods for details of the Hamiltonian). The SOC term is diagonalized by states with Wannier orbital angular momentum $l=1$ yielding at the zone center a low-energy quartet with  $j=3/2$ and a  $j=1/2$ doublet over 400 meV higher in energy.

%Just as photons can carry polarization and angular momentum\cite{Bliokh2015} electrons can carry spin and Wannier orbital angular momentum (WOAM).
%Another crucial factor is the strong trigonal warping of the Fermi pockets arising from the distinct spatial polarization of the $d_{xy}$, $d_{yz}$ and $d_{xz}$ orbitals~\cite{bruno2019band}. 
%This, combined with the strong spin-orbit coupling of Ta and inversion asymmetry of the interface, leads to a significant out-of-plane canting of the otherwise helically-polarized spin texture of the carriers in the momentum space~\cite{fu2009}. As mentioned earlier, the Rashba effect alone can enhance $H_{c2,\parallel}$ by a factor of $\sqrt{2}$. The out-of-plane spin canting, $\sigma$, can further strengthen the cooper pairs through an Ising coupling between the electrons with wave functions $\psi_{\sigma\downarrow}(k)$ and $\psi_{-\sigma}(-k)$~\cite{xi2016ising,saito2016superconductivity,xiao2012}. 
 
We can define a  Land\'e-$g$ factor  for the $t_{2g}$ derived spin-orbit coupled states as $\hat{\bm m}=-\mu_B g_L \hat{\bm j}$. 
A text-book computation~\cite{Ashcroft1976} yields $g_L=0(-2)$ for the $j=3/2(1/2)$ states, 
implying zero (finite) magnetic susceptibility. 
The trigonal crystal field further splits the $j=3/2$ quartet at the zone center into two Kramers doublets and introduces mixing between the $j=3/2$ and $j=1/2$ sectors. Additional mixing of these sectors is also provided by the kinetic energy at finite momenta. This leads to a small but finite magnetic susceptibility of the $j=3/2$ low-energy bands, certainly much smaller than the Pauli spin susceptibility of free spin-$1/2$ electrons, as shown below.

The bands emanating from the two Kramers doublets of mainly $j=3/2$ character are shown in Figure 1(b) along $\Gamma-M$ and $\Gamma-K$ ($n=1$ blue, and $n=2$ orange) and will be our main focus.
At finite momenta, their double degeneracy is split into opposite helicity branches $E_{n\pm}(\bm{k})$
by the inversion symmetry breaking Rashba term. 
%well as higher order sub-bands, over 100 meV higher in energy as seen by ARPES, and attributed to quantum confinement of electrons near the surface~\cite{bruno2019band,arnault2023}. 

Figure 1(c) shows the sixfold symmetric Fermi surface (FS) for the low- and high-chemical potential $\mu$ cases specified in Figure 1(b). Close to the zone center, both bands show a weakly hexagonal FS (Figure 1(c) bottom), and band $n=1$ develops a star-shaped FS as  $\mu$ is increased  (Figure 1(c) top). 
The magnetic moment texture $\braket{\hat{\bm{m}}(\bm{k})}$, illustrated in the same figure (only for $E_{n-}(k_F)$, one of the Rashba-split bands, for clarity), is far from a uniform magnetic moment $m_{0}=\mu_B$ of free spin-$1/2$ electrons, which can  exhibit a Landau-type orbital angular momentum for perpendicular magnetic fields but do not have a Wannier orbital angular momentum.

As expected for a small $g_L$, the modulus $m(\bm{k})$ of the low-energy spin-orbit coupled $t_{2g} $bands
is substantially anisotropic, and more importantly, it is significantly smaller than one $\mu_B$, the reduction being greater near the zone-center. This is clear in Figure 1(c), in particular for the portions of the star-shaped Fermi surface closer to the origin and the inner Fermi surface in the upper panel, and for both Fermi surfaces in the case of a reduced chemical potential in the lower panel. 
 This strong renormalization of the magnetic moment will have important consequences for the coupling with an in-plane magnetic field, substantially modifying the conventional Pauli paramagnetic response.    
%As we show in the following, this strong renormalization of the electronic spin will have important consequences for the Zeeman coupling with an in-plane magnetic field, substantially enhancing the Pauli limit.  

 Assuming superconductivity is quenched by Pauli paramagnetic pair breaking, the critical field $H_{c2,\parallel}$ of a system can be estimated by identifying the point where the normal-state and superconducting free energy densities level $f_N(H)=f_{SC}(H)$. This condition gives $H_{c2,\parallel}\sim H_P \sqrt{\chi_P/(\chi_N-\chi_{SC})}$, where $H_P=\Delta_0/\sqrt{2}$ is the Pauli field, $\chi_P=\mu_B^22N_F$ the Pauli spin susceptibility of non-interacting electrons with density of states per spin at the Fermi level $N_F$, and $\chi_{N(SC)}$ is the normal-state (superconducting) magnetic susceptibility.
The critical field $H_{c2,\parallel}$ will then exceed $H_P$ when $\chi_{SC}>0$~\cite{gor2001superconducting,yip2002} (which happens for a mixed-parity superconducting state), or when the normal state magnetic susceptibility is reduced $\chi_{N}<\chi_{P}$~\cite{yip2013,Xie2020}. Since inversion symmetry is broken in KTaO$_{3}$(111) interface, a mixed parity superconducting state with spin-singlet and spin-triplet components is allowed. However, very little is known in this system about the triplet component of the SC state that would imply a non-zero $\chi_{SC}>0$. Gor'kov and Rashba \cite{gor2001superconducting} estimated a factor of $\sqrt{2}$ enhancement from the mixed SC state in 2D Rashba-split systems, much smaller than the enhancement up to 8.4$H_P$ found in this work. 
We will therefore study and estimate the expected enhancement that arises from $\chi_{N}<\chi_{P}$ (which can then be taken as a lower bound of enhancement), and take $\chi_{SC}=0$.  

We compute the normal-state in-plane magnetic susceptibility  of band $n$ for a field along the $x\equiv [1\overline10 ]$ direction 
as 
$ \chi^{xx}_n=\mu_B^2 N_F\sum_{\eta=\pm} \gamma^x_{n\eta,F}$
where $\gamma^x_{n\eta,F}$ is the Fermi surface average of the matrix element squared, $\gamma_{n\eta}^x(\bm{k})=|\bra{n\eta\bm{k}}(-\hat l_p^x +2\hat s^x)\ket{n\eta\bm{k} }|^2 $ (see Materials and Methods for details). In the case of non-interacting electrons, i.e. without SOC nor Rashba, $\gamma_{n\pm}^x(\bm{k})=1$, and the Pauli susceptibility is recovered $\chi^{xx}_n=\chi_{P,n}=2\mu_B^2 N_{F,n}$.  
Figures 5(a),(b) show the matrix element $\gamma_{n-}^x(\bm{k})$ in the BZ for both $j=3/2$ derived low-energy bands in KTaO$_{3}$, illustrating its strong reduction from unity near the zone center and along particular directions. 
 This reduction in $\gamma_{n\eta}^x(\bm{k})$, and hence in $\gamma_{n\eta,F}^x$, implies a reduced magnetic susceptibility (much smaller than the Pauli susceptibility $\chi^{xx}_n<\chi_{P,n}$) and hence a strong enhancement of the critical field. Both the SOC and Rashba splitting contribute to this result, as SOC is responsible for the strong reduction of the Land\'e-$g$ factor and Rashba suppresses opposite-chirality mixing, which otherwise would enhance the normal state susceptibility. 

 Figure 5(c) shows the band resolved magnetic susceptibility normalized to the Pauli susceptibility $\chi_{P,n}$. It is indeed much smaller than unity for small chemical potential and grows as the Fermi surface expands and departs from the zone center. Panel (d) shows the critical field for each band separately in units of its own Pauli limit field. This very simplified model of the (111)-surface explains the large values observed in the experiment for $H_{c2, \parallel}$, regardless of which band has the dominant superconducting order parameter,  and yields even higher values at low-doping. A notable drawback is that the computation predicts the ratio  $H_{c2, \parallel}/H_P$  to decrease with doping, while in the experiment it increases. More realistic computations, however, could have more subbands over 100 meV higher in energy, as observed by ARPES, and attributed to quantum confinement of electrons near the surface~\cite{bruno2019band,arnault2023}. 
These subbands have smaller trigonal crystal field splittings~\cite{bruno2019band}, which in our computations would reflect in purer $j=3/2$ states with smaller $g_L$. The population of these subbands as doping increases may reconcile the chemical potential  dependence 
of $H_{c2, \parallel}$
with the experiment. A quantitative computation of this effect requires a multiband theory of superconductivity in KTaO$_3$ which is not available at the moment. 
A more realistic modeling should also take into account the mixing with $e_g$ states\cite{Stamokostas2018} and the possibility of a Rashba induced triplet component of the superconducting state~\cite{gor2001superconducting}, which could lead to a non-trivial dependence on the density. 

In summary, our experimental results demonstrate enhanced $H_{c2, \parallel}$ beyond the Pauli limit, which increases with carrier density. In systems with negligible spin-orbit coupling, electrons behave as spin-1/2 particles and therefore carry a $1\mu_B$ magnetic moment. We argue that the KTaO$_{3}$(111) interface hosts quasiparticles which carry a very small magnetic moment. This makes superconductivity largely immune to the Pauli paramagnetism and explains the anomalous enhancement of the in-plane critical field. KTaO$_{3}$ interfaces should be an interesting testbed for a variety of proposed theoretical models, which relate inversion symmetry broken superconductivity to spin-orbit coupling and Fermi surface spin texture \cite{ruhman2017odd, kanasugi2020ferroelectricity, yuan2021topological, dimitrova2007theory}. These results offer opportunities to enhance $H_{c2}$ by tuning the electronic structure and spin texture in crystalline 2D superconductors with strong spin-orbit coupling. 

\section{Materials and methods}

\subsection{Experimental}
Mobile charge carriers were introduced to the KTaO$_{3}$ surface in TiO$_{x}$/LaMnO$_{3}$/KTaO$_{3}$(111) heterostructures, grown in an oxide MBE chamber with base pressure of $2\times10^{-10}$ Torr. The carrier density was controlled with growth parameters, described elsewhere \cite{arnault2023, al2022superconductivity}. LaMnO$_{3}$ (10 u.c.s) was grown followed by TiO$_{x}$ (3 nm) on  KTaO$_{3}$ (111) substrate. Pure elements were evaporated from effusion cells. The elemental fluxes were calibrated using a Quartz Crystal Microbalance (QCM). 

Substrate was annealed at 600 $ ^\circ$C in oxygen partial pressure of $3\times10^{-6}$ Torr for 30 min prior to the growth. Sample was then heated to 800 $ ^\circ$C and growth started immediately to minimize the potassium loss. Oxygen partial pressure was kept at $3\times10^{-6}$ Torr during the growth of LaMnO$_{3}$. The growth was monitored by reflection high-energy electron diffraction (RHEED) which shows diffraction streaks, suggesting 2D growth of LaMnO$_{3}$ (Fig. S1). After the growth of LaMnO$_{3}$, the sample was cooled to 550 $ ^\circ$C, followed by the growth of titanium capping layer ($\sim$3 nm). The RHEED streaks disappear during the growth of titanium layer, suggesting an amorphous cap layer. The thin titanium layer oxidizes upon exposure to atmosphere and acts as a buffer, protecting the 2DEG. The growth (TiO$_{x}$/LaMnO$_{3}$) was replicated on a sapphire substrate which results in a film that is insulating ($\sim$1-10 MOhms).

Scanning transmission electron microscopy (STEM) was performed to study the film and interface structures. Cross-section samples were prepared using a Thermo Fisher Scientific Helios G4UX focused ion beam. High-angle annular dark-field (HAADF) images were obtained using a Thermo Fisher Scientific Spectra 300 X-CFEG operating at 200 kV with a convergence angle of 30 mrad and a HAADF detector with an angular range of 60-200 mrad. 

Transport measurements were carried out in van der Pauw geometry with square-shaped samples (5 mm × 5 mm). The contacts (400 nm Au) were deposited on the sample corners using a sputtering system through a shadow mask. The magneto-electric measurements (300-3 K) were carried out in a Quantum Design Physical Property Measurement System (PPMS). 

The Hall carrier density was extracted from the Hall experiments, 
$n_{2D}=-1/(eR_{H}$), where R$_{H}$ is the Hall coefficient and $e$ is the electron charge. The Hall co-efficient was calculated from a linear fit to the transverse resistance with the magnetic field ($R_{H}$=$d$R$_{xy}$/$d$B).

The sub-Kelvin magnetoelectric measurements were carried out in an Oxford dilution refrigerator with base temperature of 8 mK (50 mK with the application of high field). The lines were filtered to avoid spurious microwave frequency radiation from heating the sample. The superconductivity measurements were carried out using a lock-in amplifier (SR860) and a current source (CS580), 1 $\mu$A. The critical field was measured by sweeping the magnet (32 T Superconducting Magnet) at various temperatures and angles at NHMFL.

\subsection{Model susceptibility}
We use the same bilayer model as in Ref.~\cite{villar2021} with  inter-layer hopping $t=1$ eV, a SOC term for the $t_{2g}$ orbitals that results in a $j=3/2$-$j=1/2$ splitting of 400 meV at the zone center, a trigonal distortion that splits the $j=3/2$ quartet into two doublets separated by 15 meV [Figure 1(b)], and a Rashba hopping term $t_R=2$ meV that breaks the double degeneracy of the bands at finite momenta. 
The susceptibility is computed as %\cite{Golovina2012} 
\begin{equation}
\label{eq:chixx}
    \chi^{xx}_n=\mu_B^2\frac{\beta}{2}\sum_{\bm{k},\eta}\frac{ \gamma^x_{n\eta}(\bm{k})}{1+\cosh\left[\beta\xi_{n\eta}(\bm{k})\right]}=\mu_B^2 N_F\sum_\eta \gamma^x_{n\eta,F},
\end{equation}
where $\eta=\pm$ is the helicity index of band $n=1,2$ (split by the Rashba term), $\xi_{n\eta}(\bm{k})=E_{n\eta}(\bm{k})-\mu$ and $\beta=1/\kappa_B T$. Note that in Eq.~\eqref{eq:chixx} we are working in the usual approximation $E_{n+}(k_F)-E_{n-}(k_F)\gg \Delta$ (the maximum $\Delta\sim 0.29$ meV in Table I), in which the opposite-helicity contribution (also called inter-band contribution) to the susceptibility can be neglected, explored in Refs.~\cite{gor2001superconducting,yip2002}. 
The momentum dependent function $\gamma^x_{n\eta}(\bm{k})$ encodes the information of the electronic spin and orbital angular momentum structure of band $n$ with helicity $\eta$, and hence the magnetic moment [see Figure 1(c)], and $\gamma^x_{n\eta,F}$ is its averaged value over the Fermi surface.
Along the $x$-axis $[1\bar 1 0]$ it reads,
\begin{align}
\label{eq:gammax}
    \gamma_{n\eta}^x(\bm{k})=\sum_{\nu\nu' \sigma\sigma'}|\braket{n\eta|\nu \sigma }\bra{\nu \sigma}(-\hat l_p^x +2\hat s^x)\ket{\nu' \sigma'}\braket{\nu' \sigma'|n\eta }|^2
    %&\sum_{\mu\nu,\sigma=\pm 1/2}[u_{nl,\mu\sigma}^{*}(\bm{k})u_{nl,\mu\bar\sigma}(\bm{k})u_{nl,\nu\bar\sigma}^{*}(\bm{k})u_{nl,\nu\sigma}(\bm{k})\\\nonumber
    %&\quad+i2\sigma u_{nl,\mu\sigma}^{*}(\bm{k})u_{nl,\mu\bar\sigma}(\bm{k})u_{nl,\nu\sigma}^{*}(\bm{k})u_{nl,\nu\bar\sigma}(\bm{k})] 
\end{align}
here $\ket{n\eta}$ are the single particle states that diagonalize the Hamiltonian and 
$\ket{\nu \sigma}$ are the states in the basis spanned by 
orbital-layers labeled by $\nu$ with spin $\sigma$. $\hat l_p^x$ ($\hat s^x$) is the orbital angular momentum (spin) operator acting on the orbital (spin) part of state $\ket{\nu \sigma}$.
All states are evaluated at point $\bm{k}$ in momentum space.
The $\ket{n\eta}$ states include the effects of SOC, crystal field and inversion symmetry breaking on the electronic spin properties [Figure 1(c)]. 
In the case of non-interacting electrons, i.e. without SOC nor Rashba, $\gamma_{nl}^x(\bm{k})=1$, and the Pauli spin susceptibility is recovered $\chi^{xx}_n=\chi_{P,n}=2\mu_B^2 N_{F,n}$.

\begin{acknowledgments}
NC team was supported by the U.S. National Science Foundation under Grant No.
NSF DMR-1751455. This work made use of a Helios FIB supported by NSF (Grant No. DMR-1539918) and the Cornell Center for Materials Research (CCMR) Shared Facilities, which are supported through the NSF MRSEC Program (Grant No. DMR-1719875). A portion of this work was performed at the National High Magnetic Field Laboratory, which is supported by the National Science Foundation Cooperative Agreement No. DMR-2128556 and the State of Florida. J.L. acknowledges support from MUR, Italian Ministry for University and Research through
PRIN Projects No. 2017Z8TS5B and No. 20207ZXT4Z.
\end{acknowledgments}

\bibliography{apssamp.bib}

\begin{figure*}[htp]
    \centering
    \includegraphics[width=2\columnwidth]{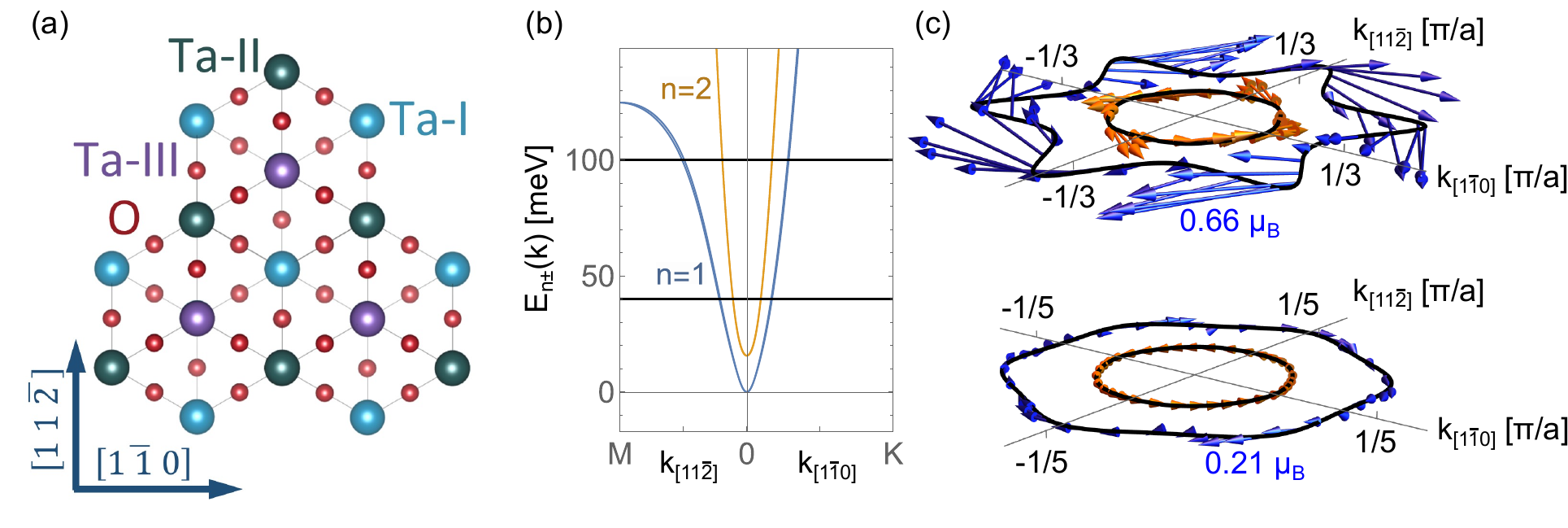}
  \caption* {\textbf{Fig. 1. Electronic structure and magnetic moment texture in a KTaO$_{3}$(111) 2D electron system.} (a) Lattice structure of the KTaO$_{3}$(111) formed by tantalum and oxygen atoms. (b) Electronic low-energy band structure of KTaO$_{3}$(111) in a two-layer model
with spin-orbit coupled $t_{2g}$ orbitals, of mainly $j=3/2$ character. The two Kramers doublets at the zone center, split by a trigonal crystal field, are further split into opposite helicity non-degenerate bands at finite $k$ by a small Rashba term (hardly visible in the plot). (c) Fermi surface and magnetic moment texture for chemical potential $\mu=40$ meV (bottom) and $\mu=100$ meV (top) specified in (b) (black lines). For clarity only the moments of the outer helicity Rashba-band $E_{n-}(k_F)$ is shown for each band: $n=1$ (blue arrows) and $n=2$ (orange arrows). }

    \label{fig:Fig1}
\end{figure*} 

\begin{figure*}[htp]
    \centering
    \includegraphics[width=2\columnwidth]{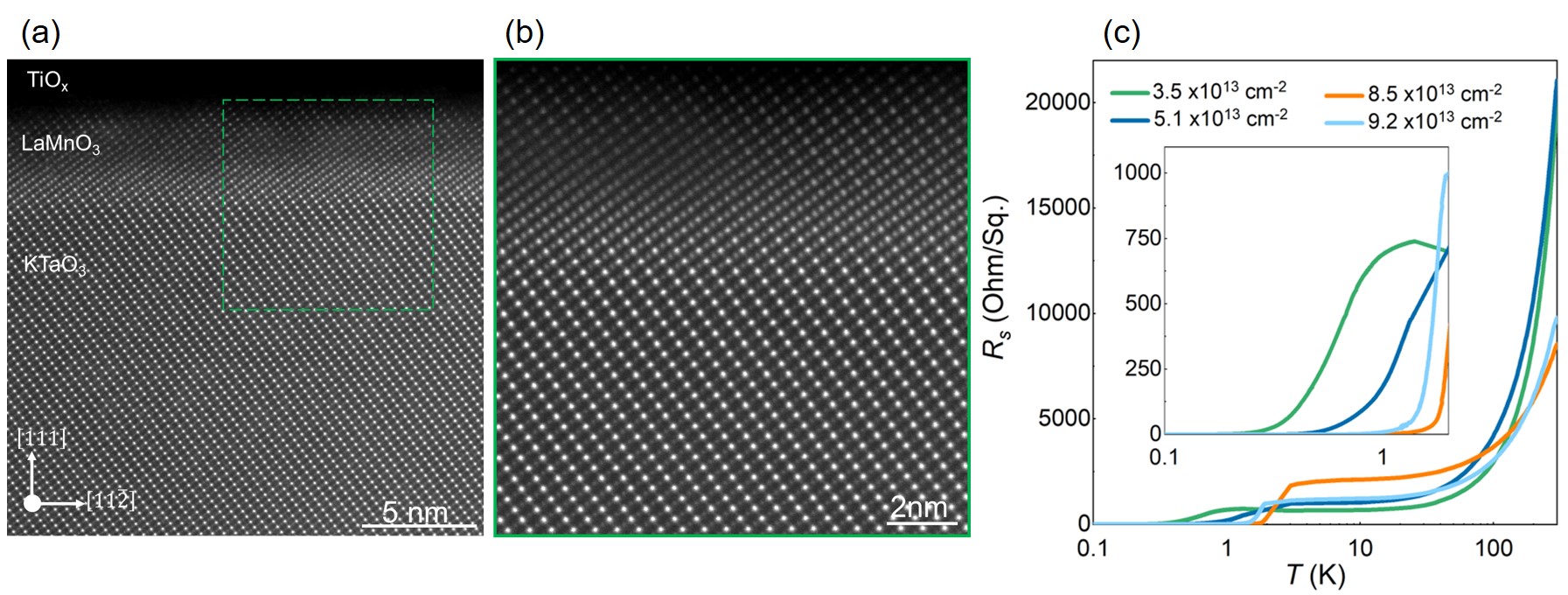}
 \caption* {\textbf{Fig. 2. Two-dimensional electron system at the epitaxial LaMnO$_{3}$/KTaO$_{3}$(111) interfaces.} (a) Cross-section HAADF-STEM images of LaMnO$_{3}$ film grown on KTaO$_{3}$ substrate. (b) Higher magnification HAADF-STEM image of LaMnO$_{3}$/KTaO$_{3}$ interface. (c) Sheet resistant with temperature (0.1-300 K), suggesting a metallic behavior, $dR/dT>0$, followed by an abrupt superconducting transition. Inset shows zoom in of low temperature sheet resistance (0.1-2 K).}
    \label{fig:Fig2}
\end{figure*} 

\begin{figure*}[htp]
    \centering
    \includegraphics[width=1.5\columnwidth]{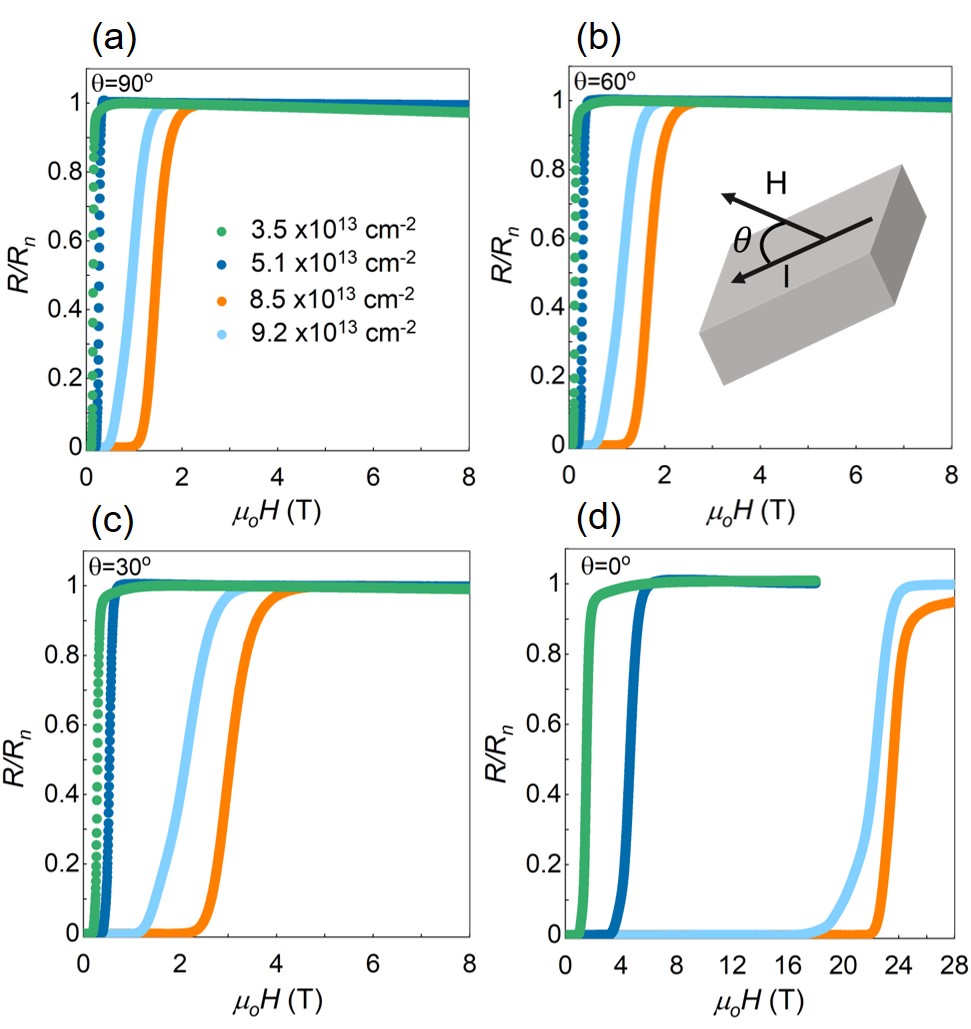}
\caption* {\textbf{Fig. 3. Angle dependence of critical field of superconductivity at the LaMnO$_{3}$/KTaO$_{3}$(111) interfaces.} The measurement is carried out at 10 mK and repeated at (a) 90$^o$, (b) 60$^o$, (c) 30$^o$, and (d) 0$^o$ between current and applied magnetic field. Cooper pairs are more resilient when the magnetic field is applied in the plane. $H_{c2}$ is sensitive to carrier density, reaching its maximum for $n=8.5\times10^{13}$ $cm^{-2}$.}
    \label{fig:Fig3}
\end{figure*} 

\begin{figure*}[htp]
    \centering
    \includegraphics[width=1.5\columnwidth]{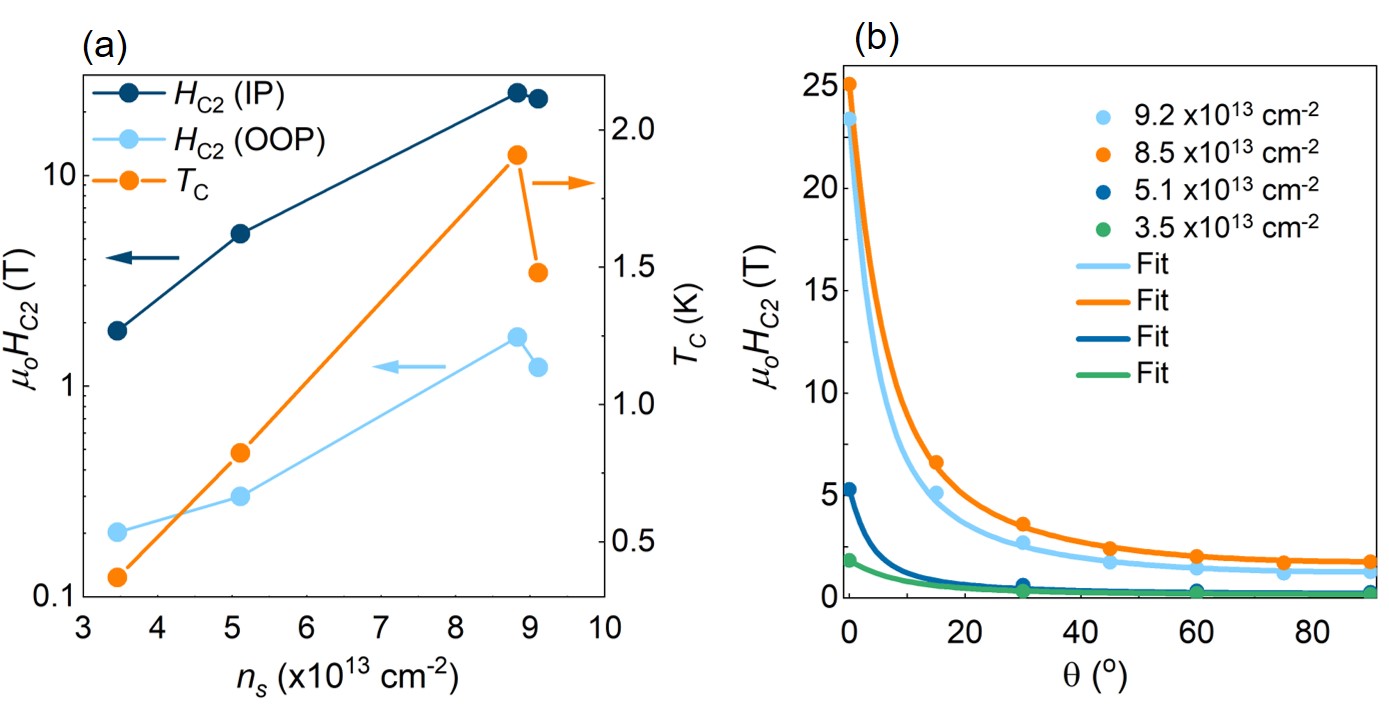}
 \caption* {\textbf{Fig. 4. Enhanced in-plane $H_{c2}$ at the epitaxial LaMnO$_{3}$/KTaO$_{3}$(111) interfaces.} (a) Superconducting transition with carrier density. $T_{c}$, $H_{c2, \parallel}$, $H_{c2, \perp}$ reach their maxima in  KTaO$_{3}$(111) 2D electron system with $8.5\times10^{13}$ $cm^{-2}$ carrier density. The critical temperature increases somewhat linearly with carrier density while critical field is much more sensitive to density of charge carriers. (b) $H_{c2}$ (defined at 0.9$R_{n}$) at various angles between current and magnetic field measured at base temperature. Equation 1 describes the results, suggesting a 2D superconductivity.}
    \label{fig:Fig4}
\end{figure*} 

\begin{figure*}[htp]
    \centering
    \includegraphics[width=2\columnwidth]{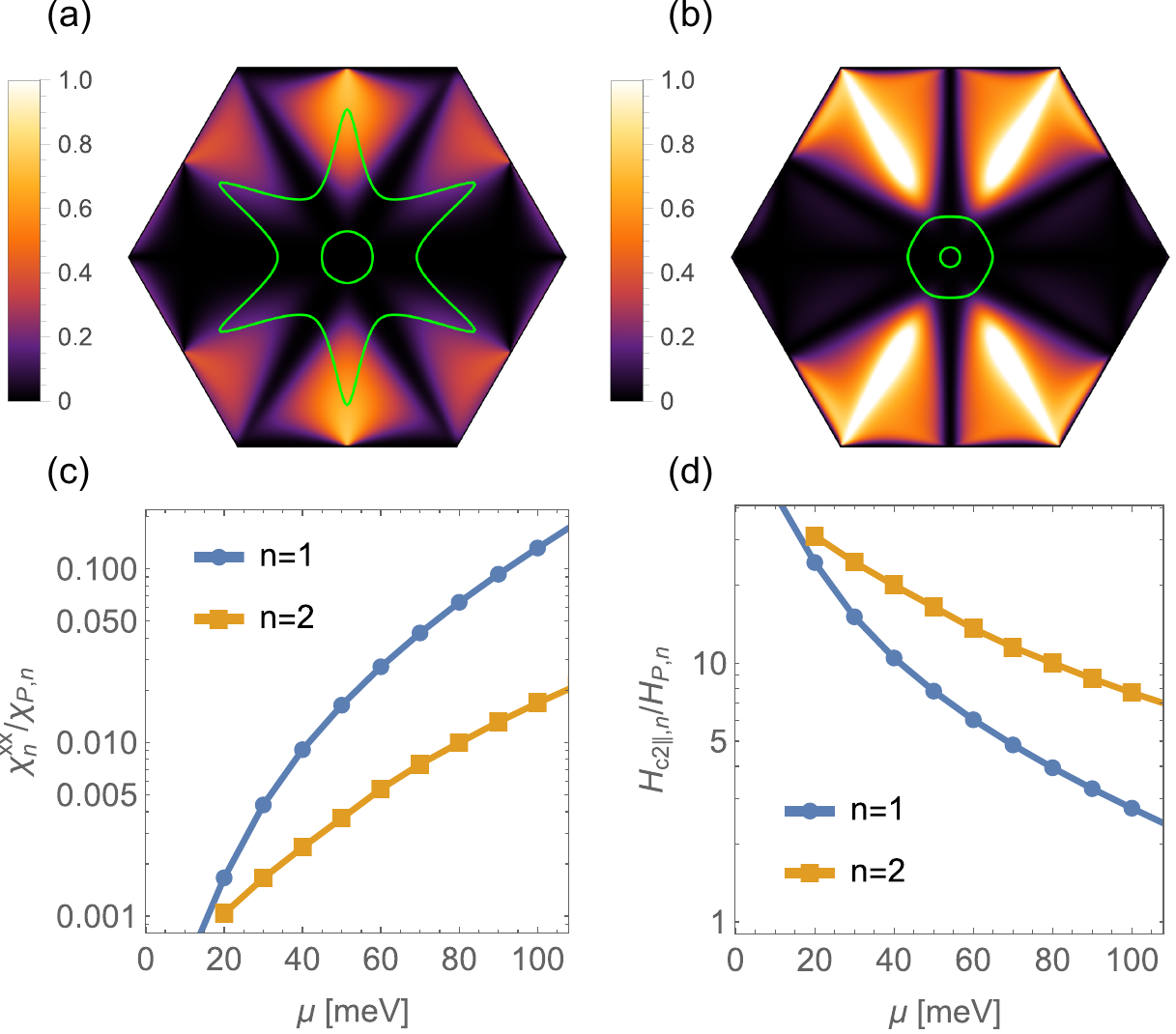}
 \caption* {\textbf{Fig. 5. Spin-orbit coupling and Rashba enhancement of $H_{c2,\parallel}$.} Matrix elements from Eq.~\eqref{eq:gammax}  for (a) $n=1$ and (b) $n=2$. The green contours show the Rashba band $E_{-}(k_F)$ Fermi surface at $\mu=40$ meV (inner contour) and $\mu=100$ meV (outer contour). (c) Spin susceptibility $\chi_{xx,n}$ of the  bands [Eq.~\eqref{eq:chixx}] normalized with the Pauli susceptibility $\chi_{P,n}$ as a function of chemical potential. (d) Corresponding $H_{c2,\parallel}$ in units of the band Pauli-limit critical field with chemical potential.}
    \label{fig:Fig5}
\end{figure*} 

% The \nocite command causes all entries in a bibliography to be printed out
% whether or not they are actually referenced in the text. This is appropriate
% for the sample file to show the different styles of references, but authors
% most likely will not want to use it.

\end{document}